\documentclass[reprint,nofootinbib,amsmath,amssymb,fontenc aps]{revtex4-1}
\usepackage{graphicx,dcolumn,bm,hyperref,float}
\usepackage{anyfontsize}
\usepackage{color}
\usepackage{xcolor}

\newcommand{\beq}{\begin{equation}}
\newcommand{\eeq}{\end{equation}}
\newcommand{\bea}{\begin{eqnarray}}
\newcommand{\eea}{\end{eqnarray}}
\newcommand{\bef}{\begin{figure}}
\newcommand{\eef}{\end{figure}}

\newcommand{\mpl}{M_{\mbox{\tiny{Pl}}}}

\newcommand{\rhode}{\rho_{\rm DE}}

\newcommand{\pde}{p_{\rm DE}}
\newcommand{\rhom}{\rho_{\rm M}}

\begin{document}

\title{Examining Quintessence Models with DESI Data}

\author{Zahra Bayat}
\email{zahra.bayat@tufts.edu}
\author{Mark P.~Hertzberg}
\email{mark.hertzberg@tufts.edu}
\affiliation{Institute of Cosmology, Department of Physics and Astronomy, Tufts University, Medford, MA 02155, USA
\looseness=-1}

\begin{abstract}
We examine data from the 
Dark Energy Spectroscopic Instrument (DESI) collaboration 
which has implications for the nature of dark energy.
We consider classes of models that manifestly obey the null energy condition, with a focus on quintessence models.
We find that hilltop potentials and exponential potentials provide modest improvement compared to a cosmological constant, but the statistical evidence is only marginal at this stage. We correct some analyses in the existing literature which attempted to compare some quintessence models to the data, giving an overly positive result.
\end{abstract}

\maketitle

\tableofcontents

\section{Introduction}
%{\em Introduction}.---

A question of fundamental importance is whether the dark energy density is evolving in time. Within the so-called $\Lambda$CDM model, the dark energy is taken to be a cosmological constant, which, by definition, does not change in time. From the point of general relativity, the inclusion of a cosmological constant is an allowed option that does not alter the number of degrees of freedom of the theory. However, if one considers more degrees of freedom than in general relativity and the Standard Model of particle physics, then one can form models for dark energy in which it evolves. The most basic option is to add a very light scalar $\varphi$ that evolves in some potential $V(\varphi)$. If the field is spatially homogeneous, and the potential energy $V(\varphi)$ dominates over its kinetic energy $\dot\varphi^2/2$, then it can act as a form of dark energy; known as ``quintessence". 

To shed light on whether dark energy is evolving, the Dark Energy Spectroscopic Instrument (DESI) collaboration has presented cosmological constraints based on Baryon Acoustic Oscillation (BAO) measurements \cite{DESI:2024mwx}. Collecting data from over 6 million extragalactic objects across redshifts with $0.295\leq z\leq2.33$, the DESI survey delivers estimations of the transverse comoving distance and the Hubble rate, normalized by the sound horizon, across seven distinct redshift intervals.

The evolution of dark energy is primarily determined by its equation of state; the ratio of pressure $P$ to energy density $\rho$
\beq
w={P\over\rho}
\eeq
If $w$ differs from $-1$, then the energy density $\rho$ depends on time. One can consider many possible parameterizations for the time dependence. 
In this work, we will follow the DESI analysis and focus on the Chevallier-Polarski-Linder (CPL) parametrization in which $w$ is taken to be linear in the scale factor
\begin{equation}
    w(a)=w_0+ (1-a)w_a
\label{CPL}\end{equation}
Here $w_0$ is today's value ($a=1$) and $w_a$ is the (negative) of the derivative today. A cosmological constant corresponds to $w_0=-1$ and $w_a=0$. Although this is a simplistic parameterization and there is no fundamental reason for it to be precise, it will turn out to be a useful way to summarize the predictions of a range of models.

The data provided in Figure 6 of Ref.~\cite{DESI:2024mwx} shows marginalized posterior constraints at $68\%$ and $95\%$ confidence from the combination of DESI BAO with CMB data and three different Supernova data sets; PantheonPlus, Union3, and DESY5. 
According to Ref.~\cite{DESI:2024mwx}, the degree of tension with the $\Lambda$CDM model (where $w_0 = -1 $ and $ w_a = 0$) is found to be $2.5\sigma$, $3.5\sigma$, and $3.9\sigma$ for each of these three datasets, respectively. This indicates the data has a fairly strong preference for evolving dark energy, especially from the latter two data sets.

If one takes the central value of the data, which is roughly around 
$w_a\sim -1$ and $w_0\sim -0.7$, then this indicates that $w$ would become $w<-1$ for $a\lesssim0.7$ (or redshift $z\gtrsim0.43)$. In such a regime with $w<-1$, the null energy condition (NEC) is violated. Since the NEC is necessary to avoid superluminality, it would be quite peculiar if it is broken by dark energy. In this work, we will focus on NEC obeying theories. We will also ask if any NEC obeying theory can provide a good fit to the data. The reason this is at least a priori plausible is that we actually do not need to take the linear CPL parameterization for all redshifts. We only need it to be approximately valid within the range of redshifts that DESI is sensitive to. However, since DESI includes redshifts well beyond $z=0.43$, one can anticipate the fit with any NEC obeying theory may not be precise.

By sticking to the standard principles of relativity ad quantum mechanics, the central possibility for evolving dark energy are quintessence models.
In fact, there has already been work in the literature on whether quintessence models can provide a good fit to the data.
In particular, we are strongly motivated by the work of Ref.~\cite{Shlivko:2024llw}, in which it is claimed that a hilltop potential provides an excellent mimic to the data (while for an exponential potential there is only modest improvement). Other relevant and interesting works include Refs.~\cite{Ramadan:2024kmn}-\cite{Cheng:2025lod}.

In this paper, we carefully study a range of quintessence models. In particular, we reexamine the analysis of Ref.~\cite{Shlivko:2024llw}. We find that the procedure used to compare the theory with data in that analysis can be highly inaccurate for some important models. This is because one must take a quintessence model and output some linear CPL parameterization for its prediction for $w(a)$. We find that while this is possible, the method used in that work is typically highly inaccurate. Instead, we find that through an improved procedure, the hilltop potential provides only modest improvement compared to a cosmological constant. We find that for highly curved hilltops, the initial field value is required to be extremely close to the hilltop and the resulting deviation from a cosmological constant is found to be negligible. We also re-examine the exponential model too. We also consider the case at the boundary of the NEC limit, in which $w$ always remains $w\ge-1$. Altogether, we find that there is only a modest improvement in the fit to data. So, for now, the $\Lambda$CDM model is still alive.

\section{Quintessence}

 \subsection{Basics}
Let us focus on quintessence models for dark energy. The most general two derivative action for a scalar $\varphi$ can be written as
\begin{equation}
    S = \int d^4 x \sqrt{-g} \left[ \frac{1}{2}  \mpl^2 R - \frac{1}{2}  g^{\mu \nu} \partial_\mu \varphi \partial_\nu \varphi - V(\varphi) \right ]  + S_m
\end{equation}
 where $\mpl=1/\sqrt{8\pi G}$ is the reduced Planck mass, $g_{\mu \nu}$ is the metric, $R$ is the Ricci curvature scalar and $S_m$ is the action for matter. We assume that in this Einstein frame, the coupling of the scalar to matter is negligible, so the theory is safe from fifth force constraints. The theory is then specified by the choice of potential function $V(\varphi)$.
 
 The scalar field’s homogeneous equation of motion is:
\begin{equation}
    \ddot{\varphi}(t)+3H(t) \dot{\varphi}(t) +\frac{dV(\varphi )}{d\varphi}=0
\label{phiode}\end{equation}
 We take the entire energy density of dark energy to be comprised of the scalar field such that there is no additional cosmological constant and the energy density is: 
\begin{equation}
    \rhode = \frac{1}{2}\dot{\varphi} (z)^2+ V(\varphi (z))
\label{rhoDE}\end{equation}
 The equation of state parameter $w$ for the scalar field dark energy is determined by:
\begin{equation}
    w=\frac{\pde}{\rhode} = \frac{\frac{1}{2}\dot{\varphi}^2- V(\varphi)}{\frac{1}{2}\dot{\varphi }^2+ V(\varphi)}
\end{equation}
which is in general a function of time (or scale factor).

\subsection{ Potentials }

The potential function $V$ determines the dynamics. Since the scalar $\varphi$ is Lorentz invariant, then so too is any potential $V(\varphi)$. This means there are a infinite number of choices allowed by relativity. So while we cannot motivate one choice over another from a low energy point of view, we can just select some classes of representative choices.

A first class of models is the exponential potential:
\begin{equation}
    V_{\rm exp}(\varphi)=V_0 \,e^{\beta \varphi/\mpl}
\end{equation}
The initial field value can be set to zero, without loss of generality (otherwise, a constant off set can be absorbed into the prefactor $V_0$). The value of the prefactor $V_0$ is assumed too be such that the total energy density inn dark energy today is the obsverved value, with $V_0\sim H_0^2 \mpl^2$. The residual parameter of the model is the dimensionless $\beta$, which controls the steepness of the exponential.

 A second class of models is the hilltop potential:
 \begin{equation}
    V_{\rm hill}(\varphi)= V_0 \left( 1 - \frac{1}{2}k^2 \varphi^2/\mpl^2 \right)
\end{equation}
In this case, the initial field value is very important. If we set $\varphi_i$ very close to zero, then the field rolls very slowly for the entire history of the universe and it behaves almost the same as a cosmological constant. While if $\varphi_i$ is larger, then the rolling is faster, and it behaves quite differently. As above, the value of $V_0\sim H_0^2\mpl^2$. Finally, the dimensionless parameter $k$ controls the curvature of the potential; for $k\sim 1$ the field rolls on the Hubble time, while if $k\gg 1$, the field rolls significantly faster; this will be an important point that we will return to.
We note that at late times the potential energy in this model will go negative, but we will not evolve to such late times, so this will not directly be a problem. In any case, one could add higher order terms, such as $+\varphi^4$, to avoid any runaway behavior at late times.

 In this work, we will numerically explore a range of values of $\beta$ and $k$ in each model. Also, we will explore a range of initial field values $\varphi_i$ for the hilltop model. We will map out the associated  $\{w_0,w_a\}$ values and perform a statistical comparison to data.

\section{Time Evolution}

 We numerically solve the equations of motion to find the evolution of both the scalar field $\varphi$
 and the scale factor $a$. 
  At late times, we only need to track matter and dark energy, so the equation of motion for the scale factor can be expressed as 
 \begin{equation}
     {\dot{a}\over a}=H={1\over\sqrt{3}\,\mpl}\sqrt{\rhom+\rhode}
 \label{Hode}\end{equation}
 where $\rhom=\rho_{\rm M,0}(a_0^3/a^3)$ is the matter density and $\rhode$ is the dark energy density given above in Eq.~(\ref{rhoDE}).
We can numerically solve the pair of differential equations (\ref{phiode}) and (\ref{Hode}), subject to initial conditions. We only need to impose that the matter density dominates over the dark energy density at early times; the precise value is not important. But what is important (at least in the case of the hilltop potential) is the initial conditions for the scalar $\varphi$. We choose an initial field velocity $\dot{\varphi}_i=0$, as the initial Hubble friction is so large the field will be frozen at early times. And, as mentioned above, for the exponential potential we choose initial condition $\varphi_i=0$ without loss of generality, and for the hilltop potential we explore a range of nonzero values $\varphi_i$.

\subsection{Results}

 \begin{figure}[t]
     \centering
     \includegraphics[width=0.94\columnwidth]{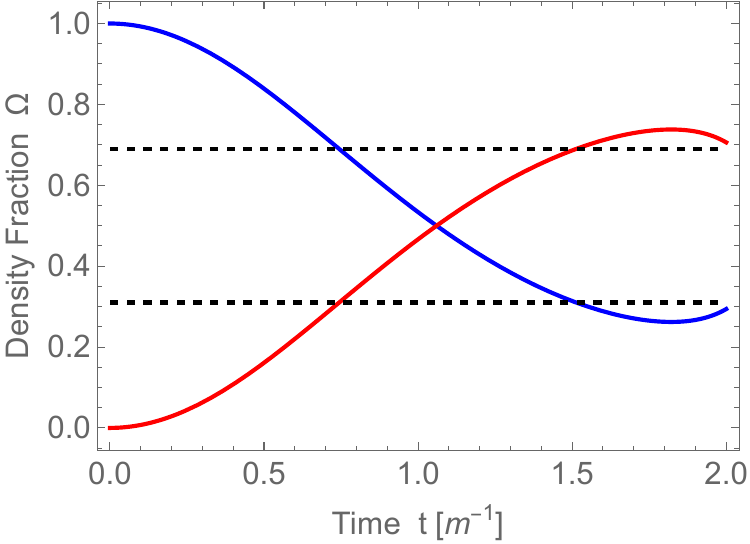}
      \includegraphics[width=0.94\columnwidth]{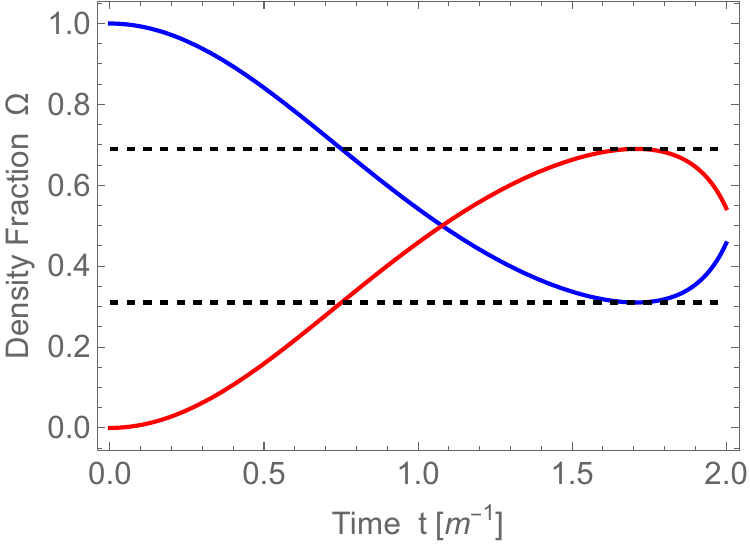}
       \includegraphics[width=0.94\columnwidth]{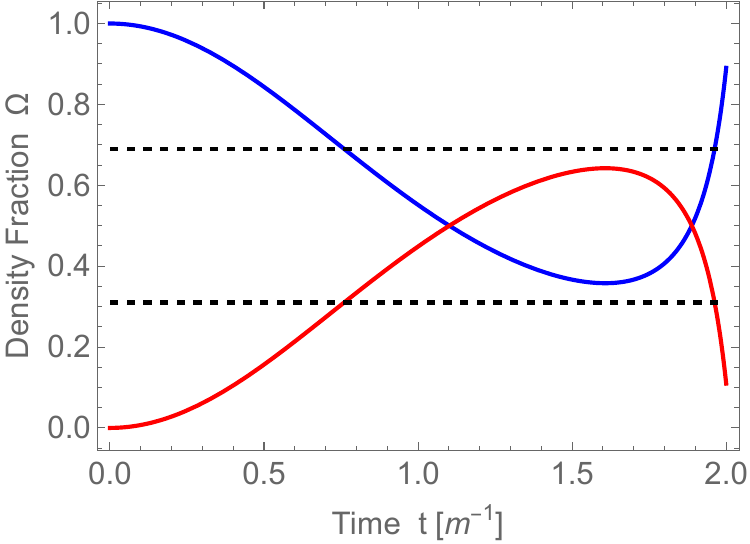}
     \caption{The evolution of the fractional energy density $\Omega$ versus time (in units of $m^{-1}$). The blue curve is for matter $\Omega_{\rm M}$. The red curve is for  dark energy $\Omega_{\rm DE}$.
     This is for hilltop potential with $k=2$. Top is $\varphi_i=0.128\mpl$, middle is $\varphi_i=0.148\mpl$, bottom is $\varphi_i=0.168\mpl$. Upper horizontal dashed line is $\Omega=0.69$; lower is $\Omega=0.31$.}
     \label{fig:ED}
 \end{figure}

We have numerically solved this system for the above potentials and different parameters ($\beta$ for exponential, and $\{k,\varphi_i\}$ for the hilltop). As an example, we output the fractional energy density versus time in Fig.~\ref{fig:ED}. The blue curve is from matter $\Omega_{\rm M}$ and the red curve is from quintessence $\Omega_{\rm DE}$. The time $t$ is given in units of $m^{-1}$, where we have written $V_0=m^2\mpl^2$. The specific value of $m$ can always be adjusted such that the duration of the universe until today is any value we want, such as $t_0=13.8$\,Gyrs.

In Standard $\Lambda$CDM cosmology, we know the values of the matter density and dark energy density quite well, namely
\begin{equation}
    \Omega_{\rm M,0}\approx0.31,\,\,\,
   \Omega_{\rm DE,0}\approx0.69
    \end{equation}
    While these could in principle change a little if one deviates from a cosmological constant, we shall assume these values are still roughly correct in this work.
We therefore impose that the present time $t_0$ is the time at which the dark energy density $\Omega_{\rm DE}$ rises up to this value. This value is indicated as the upper horizontal dashed line in the figure. Importantly, since the dark energy is dynamical here, it is not a priori guaranteed that such a moment will ever take place. In the figure we have taken the hilltop model with $k=2$. In the upper panel, $\varphi_i=0.128\mpl$, we see this moment does occur at $t\approx 1.5/m$. In the middle panel $\varphi_i=0.148\mpl$, we see that $\Omega_{\rm DE}$ just barely touches 0.69, at a time $t_0\approx1.7/m$ While in the lower panel $\varphi_i=0.168\mpl$ we see that $\Omega_{\rm DE}$ {\em never} reaches $0.69$. Therefore there is a maximum value of $\varphi_i$ in order to reach a universe today that can look like ours. For this specific case of $k=2$, we see that it is $\varphi_{i,max}\approx 0.148\mpl$. 

For the hilltop model we have explored a range of values for $k$ and report on the corresponding maximum value of $\varphi_i$ in Table \ref{tab:max3}. One can see that as the curvature factor $k$ in the potential becomes larger, the maximum allowed $\varphi_i$ (given in Planck units) becomes smaller to avoid the runaway catastrophe.
Also, in the exponential potential, while the value of $\varphi_i$ is unimportant (and can be set to zero without loss of generality), there is a maximum value of $\beta$; which is reported in the last row of Table \ref{tab:max3}.

\begin{table}[t]
    \centering
    \begin{tabular}{|l|c|c|c|}
    \hline
        \textbf{Model} & \textbf{Max $\varphi_{i}/\mpl$} 
        & \textbf{Max $w_{0}$} & \textbf{Min $w_{a}$}\\      
         %\hline
         %Exponential & - & 0.0743799 \\
         \hline
          Hilltop $k=1$ & 0.708    &-0.624 &-0.596 \\
         \hline
         Hilltop $k=1.2$ &0.502    & -0.650&-0.560\\
         \hline
         Hilltop $k=1.4$ &0.364    &-0.679 &-0.520\\
         \hline
         Hilltop $k=1.6$ & 0.267   & -0.709&-0.477\\
         \hline
         Hilltop $k=1.8$ & 0.198   & -0.730&-0.447\\
         \hline
         Hilltop $k=2$ &0.148   &-0.761 &-0.400\\
         \hline
         Hilltop $k=2.5$ & $7.23\times 10^{-2}$   & -0.816&-0.317\\
         \hline
         Hilltop $k=3$ & $3.58\times 10^{-2}$   & -0.863&-0.241\\
         \hline
         Hilltop $k=4$ & $8.91\times10^{-3}$   &-0.928  &-0.131\\
         \hline
         Hilltop $k=5$ & $2.23\times 10^{-3}$    & -0.966 &-$6.31\times10^{-2}$\\
         \hline
         Hilltop $k=6$ &$5.62\times 10^{-4}$    &  -0.977&
         -$4.48\times10^{-2}$\\
         \hline
         Hilltop $k=7$ & $1.42\times 10^{-4}$   & -0.985 &-$2.88\times10^{-2}$\\
         \hline
         Hilltop $k=8$ & $3.57\times10^{-5}$    & -0.992&-$1.47\times10^{-2}$ \\
         \hline
         Hilltop $k=9$ & $4.98\times10^{-6}$  & -0.996&-$7.33\times10^{-3}$ \\
         \hline
         Hilltop $k=10$ & $2.27\times 10^{-6}$   & -0.998&-$2.93\times10^{-3}$\\
         \hline 
          Exponential & $\beta=2.18$   & 0.0907&-0.940\\
         \hline 
     \end{tabular}
      \caption{For the hilltop model, which depends on the parameter $k$ (first column), we report on the maximum allowed value for the initial $\varphi_i$ (second column), the corresponding value of $w_0$ (third column), and the corresponding value of $w_a$ (fourth column). If $\varphi_i$ is greater than this maximum, then the field falls off the hilltop so quickly that $\Omega_{\rm DE}$ never reaches 0.69 (an example is seen in the bottom panel of Fig.~\ref{fig:ED}). For the exponential model, we report on the maximum value of $\beta$ (and the corresponding $w_0,\,w_a$) in the last row.}
      \label{tab:max3}
\end{table}

\begin{figure}[t]
     \centering
     \includegraphics[width=0.93\columnwidth]{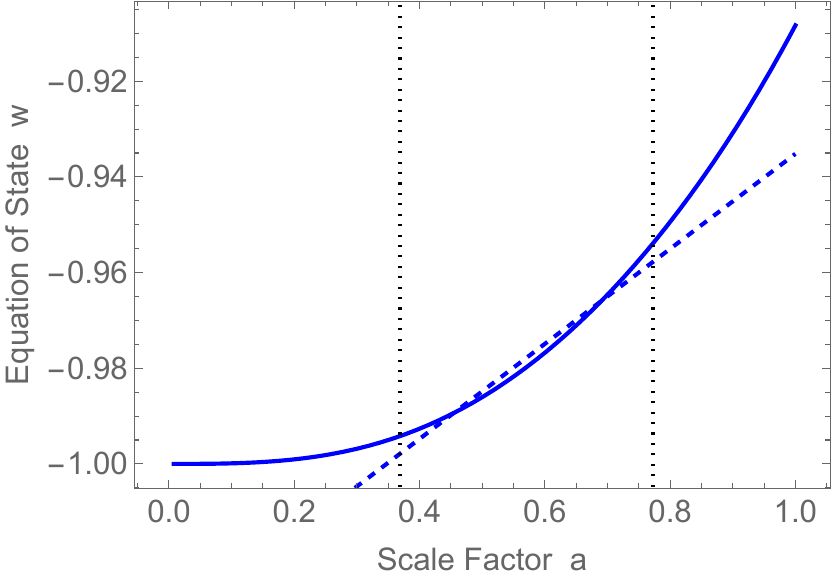}
        \includegraphics[width=0.93\columnwidth]{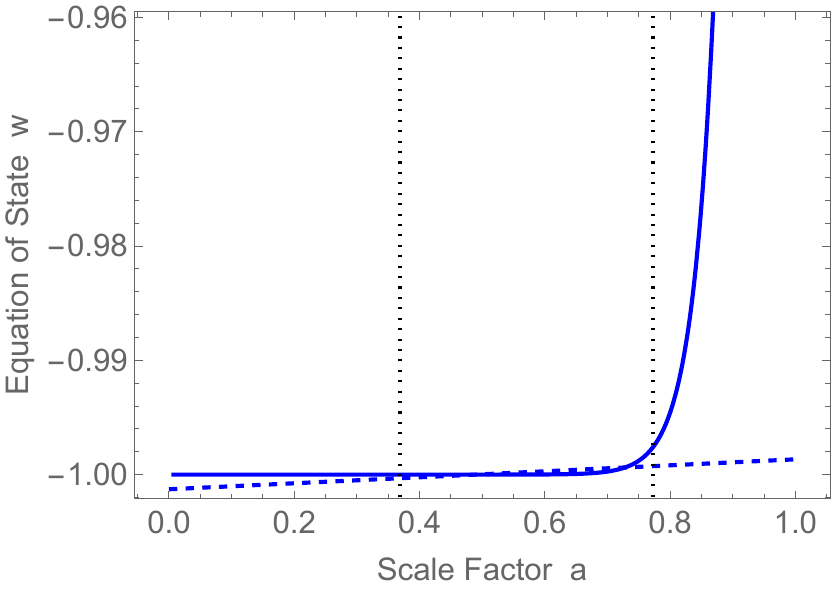} 
        \includegraphics[width=0.93\columnwidth]{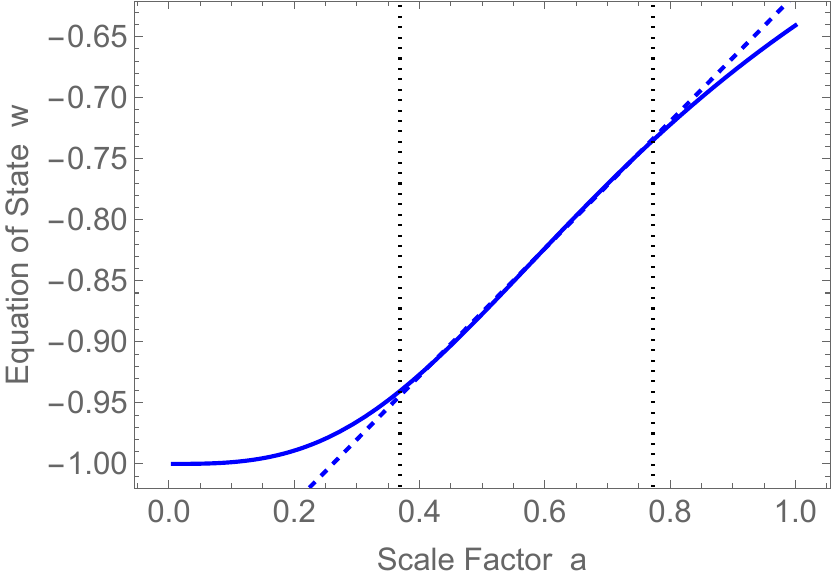}
    \caption{
    Equation of state $w$ versus scale factor $a$.
    Top panel is hilltop model with $k=1$ and $\varphi_i=0.5\,\mpl$.
    Middle panel is hilltop model with $k=10$ and $\varphi_i=2\times10^{-6}\,\mpl$.
    Bottom panel is exponential model with $\beta=1.5$.
    The solid blue curve is the exact numerical result.
    The vertical dotted lines indicate the range of scale factors that we take the DESI data to be most sensitive to.
% ; between $z=0.295$ (the lowest reliable redshift bin) up to $z=1.73$ (where dark energy would comprise $\approx10\%$ of universe in standard models). 
The blue dashed line is the best fit straight line in this interval.
 }
    \label{fig:ESa}
\end{figure}

From the evolution of $\varphi$, we can output both the energy density and the pressure and determine $w$ as a function of time or scale factor. We report on some examples of this in Figure \ref{fig:ESa}.
The evolution indicates that typically $w(a)$ is rather nonlinear; for instance $w\to-1$ in the distant past, as the quintessence field is frozen, but then evolves at later times as the field rolls. So the CPL linear parameterization for $w(a)$ does not seem accurate. However, the DESI data is only valid in a limited domain of redshift; the bins are in the range $0.295\leq z\leq 2.33$. And even for the upper bin, the data is not too sensitive to dark energy as its fractional contribution to the energy of the universe is very small even at $z\sim 2$. We choose to place a cut at $z=1.73$, which is where dark energy would comprise $\approx10\%$ of the universe in standard models. The window $0.295\leq z\leq 1.73$ is between the vertical dotted lines in the figure. Within this range, we find that a linear fit to $w(a)$ is somewhat accurate. The best linear fit in this range is given as the blue dashed line. We find that the relative integrated error between the true curve and the linear fit is typically $\lesssim1\%$ for most parameter choices of interest in this work. We then output the corresponding best fit values for $w_0$ and $w_a$ within the CPL parameterization of Eq.~(\ref{CPL}).

\begin{figure}[t]
     \centering
     \includegraphics[width=\columnwidth]{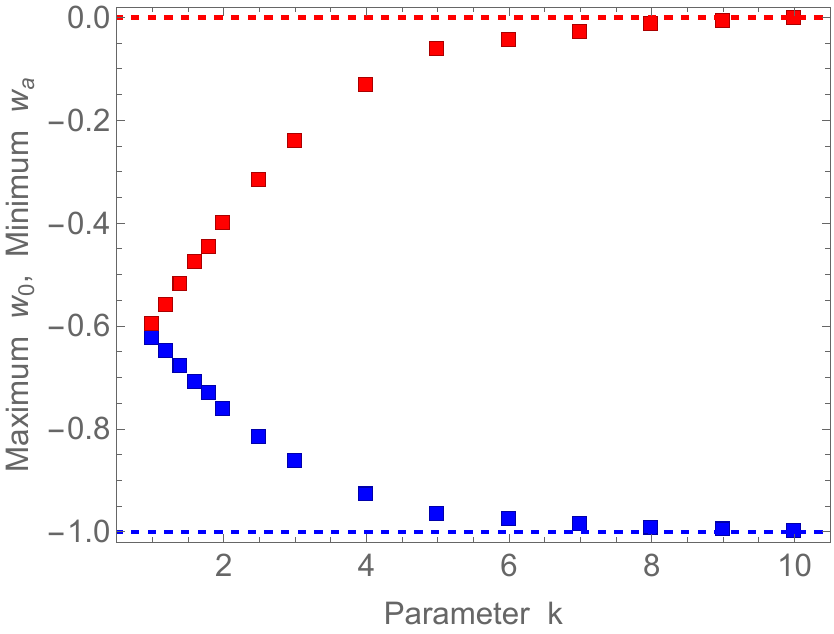}
     \caption{Blue is the maximum value of $w_0$ and red is the minimum value of $w_a$ versus parameter $k$ of the hilltop model. This visually shows the information given in Table \ref{tab:max3}. The blue dashed line is $w_0=-1$ and the red dashed line is $w_a=0$, which are the values for a cosmological constant.}
     \label{fig:MaxW0}
 \end{figure}

For the maximum value of $\varphi_i$ in the hilltop model, we show the corresponding maximum value of $w_0$ and the corresponding minimum value of $w_a$ in Figure \ref{fig:MaxW0}; this is also the data of the last two columns in Table \ref{tab:max3}.

\begin{figure}[t]
    \centering
    \includegraphics[width=\columnwidth]{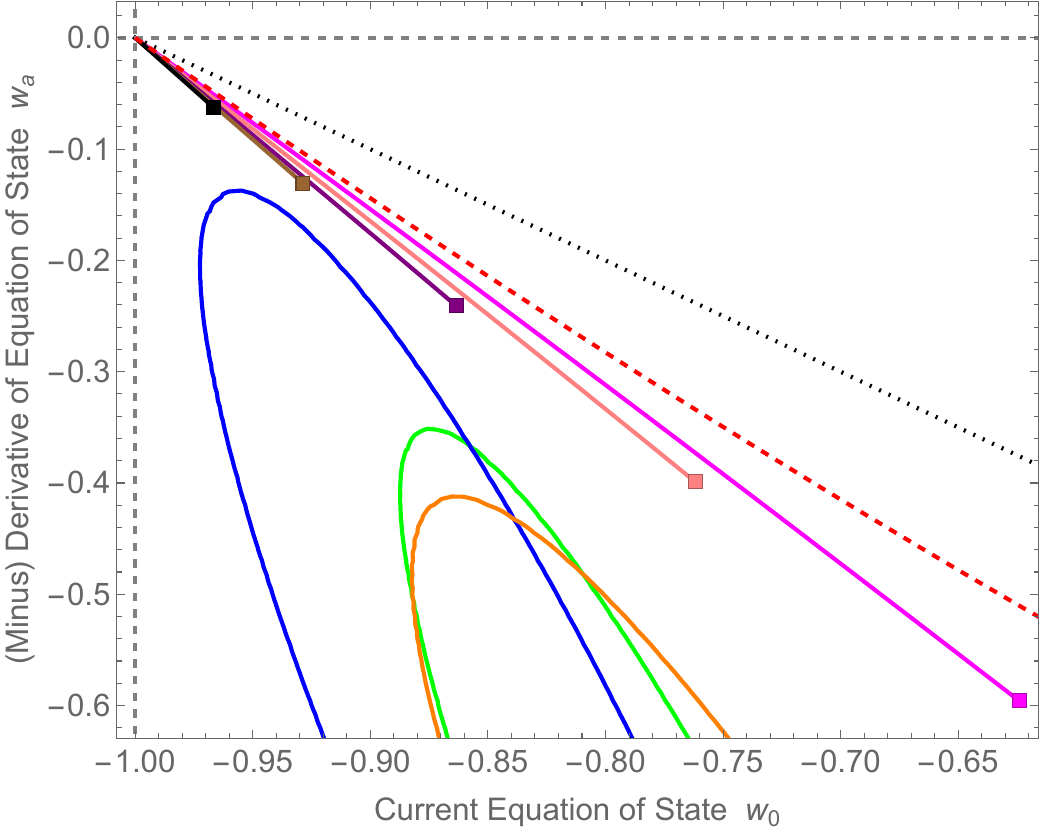}
    \caption{Equation of state parameters $w_a$ versus $w_0$ in a range of models.  
    Black dotted is NEC limit. 
    Red dashed is exponential model with varying $\beta$. 
    Magenta is hilltop model with $k=1$.
    Pink is hilltop model with $k=2$. 
    Purple is hilltop model with $k=3$.
        Brown is hilltop model with $k=4$.
    Black is hilltop model with $k=5$. 
    In these hilltop models we are varying $\varphi_i$ up to its maximum value.
    Also, the three different contours correspond approximately to the 2-sigma region of the datasets DESI BAO + CMB + PantheonPlus (blue), Union3 (orange), and DESY5 (green) from Ref.~\cite{DESI:2024mwx}.}
    \label{fig:waw0plane}
\end{figure}

For the hilltop model, for each value of $k$, we then explore a range of values $0<\varphi_i<\varphi_{i,max}$, while for the exponential model, we explore a range of values $0<\beta<\beta_{max}$. For each value, we output the corresponding best-fit values for $w_0$ and $w_a$. 
This gives the full range of allowed values in the $\{w_0,w_a\}$ plane; our results are given in Figure \ref{fig:waw0plane}.
In the figure we give several examples of the $k$ value for the hilltop model, as well as the exponential model. 
We have also indicated the 2-sigma contours of the DESI analysis; PantheonPlus (cyan), Union3 (orange), DESY5 (green).

We see that as we increase $k$, the entire space of solutions gets pushed closer and closer to the upper left corner approaching $w_0\to -1$ and $w_a\to 0$; which are the values for a cosmological constant. So, although a highly curved hilltop potential can have properties very different from a constant energy density, this does not manifest itself over the limited range of redshift that DESI is primarily sensitive. One can see this in the middle panel of Figure \ref{fig:ESa}, which is the case of $k=10$. Although the equation of state does ultimately deviate from $w=-1$ appreciably at late times, it only does so at very low redshift. At more moderate redshifts, $z\gtrsim 0.5$, it remains extremely close to $w=-1$, as the field needs to be perched very close to the hilltop initially or else it would have rolled off far too quickly and never obtained $\Omega_{\rm DE}=0.69$, as we mentioned above. 

We note that this finding is in stark contrast to Ref.~\cite{Shlivko:2024llw}, where it is claimed that the $k=10$ hilltop model can appear to fit right through the center of the DESI data. However, we find that this is a result of Ref.~\cite{Shlivko:2024llw} performing a highly inaccurate linearization of the $w(a)$ function. Instead the more correct linearization procedure leads to the high $k$ value moving to the upper left corner in the $\{w_0,w_a\}$ plane, which does not explain the DESI data.

We have also drawn the curve (black dashed line) that we refer to as the NEC limit. This is defined as the case in which $w_0$ and $w_a$ are related by:
\begin{equation}
    w_a=-1-w_0\,\,\,\,(\mbox{NEC limit)}
\label{waNECL}\end{equation}
With this choice, the relation between $w$ and $a$ is 
\begin{equation}
    w(a)=-1+a+a\,w_0\,\,\,\,(\mbox{NEC limit)}
\end{equation}
This has the property that if we now take the CPL parameterization seriously all the way to $a\to 0$, we obtain $w\to -1$. So it never violates the NEC, but is right at the limit in the early universe.

\section{Statistical Analysis}

The DESI data provides the confidence intervals in the $\{w_0,w_a\}$ plane. The contours of fixed probability density are approximately ellipses. The distribution we take to be Gaussian, i.e., the joint p.d.f is of the form
\begin{eqnarray}
&&P(w_0,w_a)\nonumber\\
&&\propto
\exp\left(-c_1w_0^2-c_2w_a^2-c_3w_0w_a-c_4w_0-c_5w_a\right)\,\,\,
\,\,\,\,\,\,\,\,
\end{eqnarray}
where the prefactors $c_1,\ldots,c_5$ are properties of the particular data set used in the analysis (PantheonPlus or Union3 or DESY5).

Then within each model, one can express $w_a$ as a function of $w_0$; some examples of this are seen in Figure \ref{fig:waw0plane}. 
This reduces the probability density to a single variable density
\begin{equation}
    p(w_0)\propto P(w_0,w_a(w_0))
\end{equation}
In the case of the hilltop model, this means fixing $k$ and allowing $\varphi_i$ to vary from 0 to $\varphi_{i,max}$. In the case of exponential model, this means varying $\beta$ from 0 to $\beta_{max}$. 

\begin{figure}[t]
     \centering
     \includegraphics[width=\columnwidth]{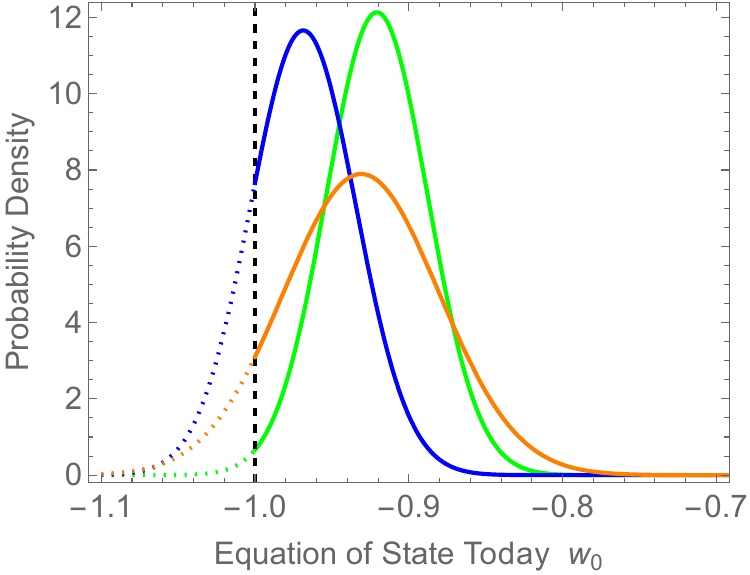}
      \includegraphics[width=\columnwidth]{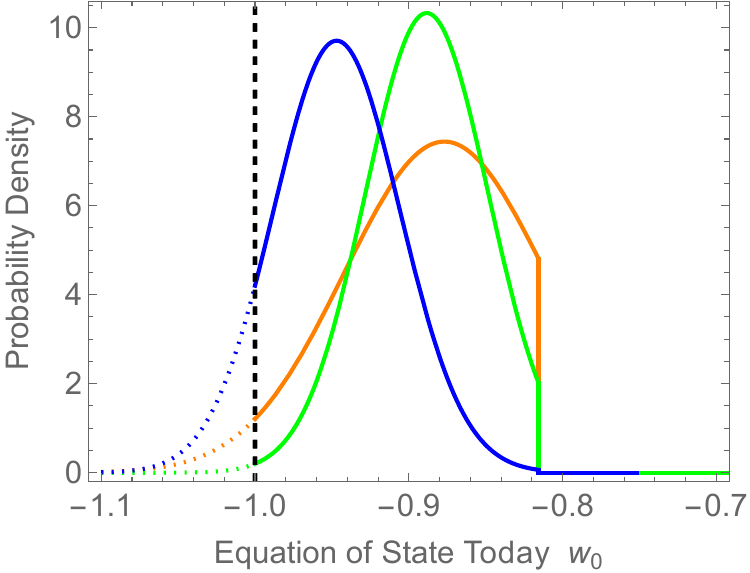}
     \caption{Probability density versus equation of state today $w_0$.
     The top panel is the NEC limit.
     The bottom panel is hilltop model with $k=2.5$.
     The blue is PantheonPlus, orange is Union3, and green is DESY5.
     The vertical dashed line is the cosmological constant value of $w_0=-1$. In the lower panel, the distribution is cut at $w_{\rm 0,max}=-0.816$. 
     For $w_0<-1$, we have changed the curves from a solid line to a dotted line; this is to indicate that this regime is an extrapolation for the purpose of a statistical analysis.}
     \label{fig:pNEC}
 \end{figure}

In the case of the NEC limit, this simply means imposing the constraint in Eq.~(\ref{waNECL}). 
This leads to the probability density given in the upper panel of Figure \ref{fig:pNEC}.

In most of the cases, we previously saw that $w_0$ has a maximum value $w_{\rm 0,max}$; so we impose the constraint $w_0\leq w_{\rm 0,max}$ when we use these distributions and normalize accordingly.
As an example, for the hilltop model with $k=2.5$ we show this probability density in the lower panel of Figure \ref{fig:pNEC}. We see that at the maximum value $w_{\rm 0,max}=-0.816$, we cut off the distribution by setting it to zero for $w_0>w_{\rm 0,max}$.

Now we wish to determine whether the data disfavors a cosmological constant ($w_0=-1$), given these probability densities.
Since we only consider $w\ge-1$ to be physical (given the constraint from the NEC), it is difficult to have a direct test of $w=-1$ as it is at the boundary of the physical domain. Nevertheless we choose to consider the following statistic: we effectively extend the probability distribution to values lower than $w_0=-1$ (through a simple interpolation) and assign a probability to the null hypothesis $w_0=-1$ of
\begin{equation}
p_{\rm null}=
{1\over\mathcal{N}}\int_{-\infty}^{-1}dw_0\,p(w_0)
\end{equation}
where the normalization constant 
\begin{equation}
\mathcal{N}=    
\int_{-\infty}^{w_{\rm 0,max}}dw_0\,p(w_0)
\end{equation}
makes it manifest that we truncate at the maximum value $w_{\rm 0,max}$. Note that here we can refer to $w_0=-1$ or $w=-1$ interchangeably, because in all these models $w_a=0$ when $w_0=-1$.

We have carried out this statistical analysis and report on our results in Table \ref{tab:stat}. In the first set of rows we report on the results for the hilltop model for different values of $k$. Then we report on the results for the exponential model. Finally, we report on the results for the the NEC limit.
In the columns we give the results for the three differet data sets.
For each we report on 2 statistics: 
(i) we report on $p_{\rm null}$; the probability of the cosmological constant null hypothesis, and 
(ii) the corresponding number of standard deviations $N_\sigma$. Since the truncated distribution is not quite Gaussian, the counting of standard deviations is non-standard. We only use it as a (non-essential) way of converting probabilities to standard notation. It is defined by
\begin{equation}
%p_{\rm null}={1\over2}-{1\over2}\mbox{Erf}\left[N_\sigma\over\sqrt{2}\right]
N_\sigma=\sqrt{2}\,\mbox{Erf}^{-1}[1-2\,p_{\rm null}]
\end{equation}
where $\mbox{Erf}^{-1}$ is the inverse of the error function.
In the table, when $p_{\rm null}>0.5$ we do not report on its precise value or the number of standard deviations, as it is not useful; in these cases there is no tension whatsoever.

\begin{table*}[h]
    \centering
    \begin{tabular}{|l|c|c|c|c|c|c|}
    \hline
        \textbf{Model}  & \textbf{Pan-Plus $p_{\rm null}$} &\textbf{Pan-Plus $N_\sigma$} &
        \textbf{Union3 $p_{\rm null}$} &   \textbf{Union3 $N_\sigma$} & \textbf{DESY5 $p_{\rm null}$} &\textbf{DESY5 $N_\sigma$}\\
         \hline
          Hilltop $k=1$  &0.115& 1.20 & 0.038&1.77 &0.0034& 2.71\\
         \hline
         Hilltop $k=1.2$  &0.113& 1.21 & 0.037&1.79 &0.0033& 2.72\\
         \hline
         Hilltop $k=1.4$ &0.111& 1.22 & 0.036&1.80 &0.0031& 2.73 \\
         \hline
         Hilltop $k=1.6$  &0.108& 1.24 & 0.034&1.82&0.0030& 2.75 \\
         \hline
         Hilltop $k=1.8$  &0.106& 1.25 & 0.033&1.84 &0.0029& 2.76\\
         \hline
         Hilltop $k=2$  &0.103& 1.26 & 0.031&1.85&0.0027& 2.77 \\
         \hline
          Hilltop $k=2.5$  &0.095& 1.31 & 0.034&1.83 &0.0025& 2.80\\
         \hline
         Hilltop $k=3$  &0.093& 1.32 & 0.045&1.69&0.0031& 2.73 \\
         \hline
         Hilltop $k=4$  &0.133& 1.11 & 0.125&1.15&0.0151& 2.17 \\
         \hline
         Hilltop $k=5$  &0.297& 0.53 & 0.33&0.43 &0.10& 1.28\\
         \hline
         Hilltop $k=6$ &$0.41$& 0.24 & $0.45$&0.12 &0.189& 0.88 \\
         \hline
         Hilltop $k=7$  &$>0.5$& - & $>0.5$&- &0.37& 0.42\\
         \hline
         Hilltop $k=8$ &$>0.5$& - & $>0.5$&- &$>0.5$& - \\
         \hline
         Hilltop $k=9$  &$>0.5$& - & $>0.5$&- &$>0.5$& -\\
         \hline
         Hilltop $k=10$  &$>0.5$& - & $>0.5$&- &$>0.5$& -\\ 
         \hline
         Exponential  &0.126& 1.15 & 0.047&1.67 &0.0041& 2.64\\
         \hline
          NEC limit   & 0.180&0.917&0.086 & 1.37 & 0.0079& 2.41\\
         \hline
    \end{tabular}
      \caption{
     Statistical analysis of each model. The columns represent the results for the 3 different data sets; they all include DESI BAO + CMB, plus the third data set indicated (PantheonPlus, Union3, and DESY5). The columns labeled $p_{\rm null}$ give the probability of the null hypothesis (cosmological constant) and the columns labeled $N_\sigma$ give the corresponding number of standard deviations. 
      The first set of rows gives the hilltop model for different values of $k$. 
      The second last row gives the exponential model. 
      The last row gives the NEC limit.
      When the probabilities are greater than 0.5, we do not provide the number of standard deviations as it is not meaningful.}
      \label{tab:stat}
\end{table*}

We see that in the exponential model the number of standard deviations is $1.15,\,1.67,\,2.64$, respectively, for the 3 data sets. While for the NEC limit the number of standard deviations is $0.917,\,1.37,\,2.41$. We note that in both of these cases, the level of tension is quite reduced compared to the $2.5,\,3.5,\,3.9$ reported by DESI in the full $\{w_0,w_a\}$ plane.

Finally, we have a range of results for the hilltop model for different values of $k$. We see that for small $k$ there is reduced tension and for large $k$ there is very little tension; the regime of the largest tension is around $k\sim2$ or so. For example, for $k=2$, the number of standard deviations is $1.26,\,1.85,\,2.77$. Again these are reduced compared to those report by DESI in the full $\{w_0,w_a\}$ plane. We note that since these models involve a new degree of freedom and a parameter $k$, these statistics should have a further penalty. Overall there is at best very modest preference for these quintessence models over a cosmological constant.

Importantly, we see that for sufficiently large $k$ our above statistic has $p_{\rm null}>0.5$, meaning that in such a regime there is no tension whatsoever. 
  This includes the case $k=10$, which was claimed in Ref.~\cite{Shlivko:2024llw} to provide a great mimic to the data and large improvement over a cosmological constant. Here we see that such a model is in fact statistically indistinguishable from a cosmological constant. Since it involves more parameters and degrees of freedom, there is no preference for this large $k=10$ case.

\section{Outlook}

In this work we have considered the initial DESI data and whether it has  preference for evolving dark energy in a range of quintessence models. We find that there is currently only modest improvement, so the cosmological constant is still viable, for now.

Important future work is to also include the updated DESI data \cite{DESI:2025zgx}. We think it is useful to restrict such analyses to models that obey the null energy condition, although it is intriguing to check to what extent this may be in tension with data. We do not think there is currently the extraordinary evidence to abandon the NEC, which can lead to problems with causality, but one can have an open mind.

Furthermore, while we have explored three classes of theories (quintessence with a hilltop potential, quintessence with an exponential potential, and also the NEC limit), it is important to explore more general possibilities. If one imposes the constraint from the NEC, one cannot deviate too strongly from these types of models, but it is conceivable some other type of potential function could appreciably improve the fit. However, models that have many features to provide a fit, also require a statistical penalty from each additional parameter.

\section*{Acknowledgments}
%{\em Acknowledgments}.---
M.~P.~H.~ is supported in part by National Science Foundation grant PHY-2310572.

\end{document}